\pdfoutput=1
\documentclass[aps,prl,reprint,twocolumn,groupedaddress,showpacs]{revtex4-1}
\usepackage[utf8]{inputenc}
\usepackage[T1]{fontenc}

\usepackage{braket}
\usepackage{amsmath}
\usepackage{amssymb}
\usepackage{csquotes}
\usepackage{bm}
\usepackage{graphicx}

\def\BZ{\ensuremath{\text{BZ}}}
\def\WZ{\ensuremath{\text{WZ}}}
\def\dd{\ensuremath{\mathrm{d}}}
\def\ee{\ensuremath{\mathrm{e}}}
\def\ii{\ensuremath{\mathrm{i}}}
\def\TR{\ensuremath{\Theta}}
\def\CC{\ensuremath{C}}

\def\tr{\ensuremath{\operatorname{tr}}}
\def\deg{\ensuremath{\operatorname{deg}}}

\usepackage[svgnames,x11names]{xcolor}

\usepackage{hyperref}
\usepackage{bookmark}

\begin{document}

\title{Topological Index for Periodically Driven Time-Reversal Invariant 2D Systems}

\author{David Carpentier}
\author{Pierre Delplace}
\email[Corresponding author~:~]{pierre.delplace@ens-lyon.fr}
\author{Michel Fruchart}
\author{Krzysztof Gawędzki}

\affiliation{Laboratoire de Physique, École Normale Supérieure de Lyon, 47 allée d'Italie, 69007 Lyon, France}

\date{\today}

\begin{abstract}
We define a new $\bm{Z}_2$-valued index to characterize the topological properties of periodically driven two dimensional crystals
when the time-reversal symmetry is enforced. This index is associated with a spectral gap of the evolution operator over one period of time.
When two such gaps are present,  the Kane-Mele index of the eigenstates with eigenvalues between the gaps is recovered as the difference of the gap indices.
This leads to an expression for the Kane-Mele invariant in terms of the Wess-Zumino amplitude.
We illustrate the relation of the new index to the edge states in finite geometries by numerically solving an explicit model on the square lattice 
that is periodically driven in a time-reversal invariant way.
\end{abstract}


\maketitle

\vspace*{-0.75cm}

\textit{Introduction.}~-- 
The recent discovery of the quantum spin Hall effect~\cite{KaneMele2005a,KaneMele2005,BernevigHughesZhang2006,Konig2007} renewed interest in topological insulating phases 
which were first encountered in the beginning of the 1980s with the 
discovery of the quantum Hall effect (QHE) \cite{KlitzingDordaPepper1980,ThoulessKohmotoNightingaleNijs1982,AvronSeilerSimon1983,Bellissard1994}.
In his seminal paper \cite{Laughlin1981} of 1981, Laughlin related the quantized Hall conductance to a quantum pump adiabatically driven by the magnetic flux.
As shown by Thouless \cite{Thouless1983}, such pumps drive through the insulator an integer number of charges whose origin is topological. 
In deep analogy, several works interpreted topological insulators and their robust boundary states in terms of
quantum adiabatic pumps \cite{FuKane2007,MeidanMicklitzBrouwer2011,FulgaHasslerAkhmerov2012}.
Interestingly, quantum crystals can exhibit original topological features when periodically driven \textit{beyond} the adiabatic regime.
While such modulation was first proposed to trigger a topological phase transition \cite{Inoue10,LindnerRefaelGalitski2011,Kitagawa11}, 
it can also yield specific topological properties which cannot be understood within the usual framework of topological band theory \cite{KitagawaBergRudnerDemler2010,RudnerLindnerBergLevin2013}.
The search for these so-called \textit{Floquet} topological states quickly became a very active field 
and has recently stimulated numerous
experimental works.
A realization of such phases in condensed matter is quite challenging \cite{Wang13,Onishi14}. 
Several alternative artificial systems have been proposed to simulate and probe analogous phases, such as 
lattices of photonic resonators periodically driven by electro-optic modulators \cite{Fang12}, ring resonator lattices \cite{Pasek14}, or more recently, photons coupled to excitons in semiconductors \cite{Karzig14}.
Signatures of topological Floquet states have already been revealed in one-dimensional quantum walks with photons \cite{Kitagawa2012}, as well as in 2D waveguide lattices \cite{Rechtsman2013}.
Shaken trapped cold atoms were also proposed as a good candidate \cite{Hauke12,Zheng14,Reichl14} and non-trivial topological phases were recently observed there~\cite{Jotzu14}.

Remarkably, although dissipation is inherent to driven systems, signatures of the topological properties observed in these experiments
can be captured by hermitian or unitary operators. In particular, topological 
properties of 2D periodically driven systems with no additional
symmetry are well characterized by the invariant proposed by Rudner \textit{et al.} \cite{RudnerLindnerBergLevin2013}
generalizing the description of static band insulators in terms of first Chern numbers.
For equilibrium phases
in the presence of symmetries, indices different from Chern numbers are required to describe topological properties.
An analogous treatment for periodically driven systems is still lacking.
Different works addressed recently this question in 1D systems \cite{Jiang11,Asboth12,Tarasinski14}.
Of great importance is, however, the case of 2D periodically-driven time-reversal invariant (TRI) fermionic systems.
In this Letter we introduce a novel topological index that generalizes the Kane-Mele invariant for the equilibrium quantum spin Hall phases~\cite{KaneMele2005}
in the spirit of the invariant of \cite{RudnerLindnerBergLevin2013} for unconstrained periodically driven systems. 

We define a $\bm{Z}_2$-valued quantity $K_{\epsilon}[U]$ that 
depends on the (quasienergy) spectral
gap $\epsilon$ of the evolution operator 
over one period of time~$U(T)$ in such a way that its difference for two distinct quasienergy gaps~$\epsilon$ and~$\epsilon'$ satisfies the relation  
\begin{equation}
	\text{KM}(\mathcal{E}_{\epsilon,\epsilon'}) = K_{\epsilon'}[U] 
- K_{\epsilon}[U],
	\label{eq:KMK}
\end{equation}
where $\text{KM}(\mathcal{E}_{\epsilon,\epsilon'})$
is the Kane-Mele invariant of the vector bundle $\mathcal{E}_{\epsilon,\epsilon'}$ of eigenstates of $U(T)$ associated to the quasienergy band between the two gaps.
Moreover, we find that this index builds unexpected bridges between the Kane-Mele invariant and the 
2D Wess-Zumino action functional, 
providing a new expression for the static Kane-Mele index that brings the field theory toolbox
to its analysis. 

In the following, we first review the construction of the index~$W$ of \cite{RudnerLindnerBergLevin2013} for the case with no symmetry. 
Then, we define the new invariant for TRI driven crystals.
We finally illustrate some topological properties of the TRI periodically driven systems on a simple lattice model.

\textit{Invariant for 2D periodically driven systems.}~--
The Hamiltonian $H$ describing a system on a translation invariant lattice may be block-diagonalized by 
the Fourier transform. This produces the Bloch Hamiltonians $H(k)$ acting on $\bm{C}^{N}$ 
for $k$ in the Brillouin torus ($\BZ$), where~$N$ is the (finite) number of internal degrees of freedom.
It is always possible to assure that $H(k)=H(k+G)$ for~$G$ in the reciprocal lattice.
A periodically driven system may be described by a time
periodic Hamiltonian which corresponds to Bloch Hamiltonians periodic both in quasi-momenta and in time, $H(t,k)=H(t,k+G)=H(t+T,k)$.

The time evolution of such systems is described by unitary operators $U(t,k) \in U(N)$ satisfying the equation $\ii \dot{U}(t,k) = H(t,k) \, U(t,k)$
with initial condition $U(0,k)=I$. Operators $U(t,k)$ define a smooth
mapping from $[0,T]\times\BZ$ to $U(N)$.
A natural invariant that characterizes the topological properties of smooth
maps between two manifolds is their homotopy class \cite{DubrovinFomenkoNovikovII},
which is however trivial for $U$.
This could be different if $U$ were periodic in time defining
a map from $S^1\times\BZ$ to $U(N)$. One may periodize $U$ in a natural way using Floquet theory
if unitary operators $U(T,k)$ have a common spectral gap~\cite{RudnerLindnerBergLevin2013}.

To do so explicitly, one diagonalizes the unitary operator $U(T,k)$ 
(this is the essence of the Floquet theory) as
\begin{equation}
	U(T,k) = \sum_{j=1}^{N} \lambda_j \, \ket{\psi_j}\!\bra{\psi_j}
\end{equation}
and one defines the effective Hamiltonian
\begin{equation}
	H^{\text{eff}}_{\epsilon}(k) = \frac{\ii}{T} \sum_{j=1}^{N} \ln_{-T \epsilon} (\lambda_j) \, \ket{\psi_j}\!\bra{\psi_j},
\end{equation}
where $\ln_{T \epsilon}$ is the logarithm with cut at argument $-T \epsilon$, so that $U(T,k) = 
\ee^{-\ii T H^{\text{eff}}_\epsilon(k)}$. 
For this effective Hamiltonian to depend smoothly on $k$, $\ee^{-\ii T \epsilon}$ has to lie in an eigenvalue gap of $U(T,k)$ for all $k$.
The quantities $\epsilon_j$ such that $\lambda_j=\ee^{-\ii T \epsilon_j}$
are called \emph{quasienergies}, so this gap is a quasienergy gap.
This allows to define the periodized versions of $U(t,k)$
\begin{equation}
	V_{\epsilon}(t,k) = U(t,k) \, \ee^{\ii t H^{\text{eff}}_\epsilon(k)}
\end{equation}
which satisfy $V_{\epsilon}(0,k)=I=V_{\epsilon}(T,k)$.
The maps $V_\epsilon$, explicitly dependent on the cut $- T \epsilon$,
may be considered as defined on the 3-torus $S^1\times\BZ$.
As described in \cite{RudnerLindnerBergLevin2013}, 
the integer-valued integral \cite{BottSeeley1978}
\begin{equation}
W_{\epsilon}[U] = \frac{1}{24\pi^2}
\mkern-9mu \int\limits_{[0,T]\times\BZ} \mkern-9mu 
\tr(V_\epsilon^{-1}\dd V_\epsilon)^3\equiv\deg(V_{\epsilon})\,,
\end{equation}
which we shall, somewhat abusively, call the degree of map~$V_\epsilon$,
defines a topological invariant that 
may be associated to the gap containing $\epsilon$.
It is well defined until the gap closes. 
The invariants $W$ are thus attached to gaps in the spectrum unlike 
the first Chern numbers that are attached to spectral bands.
Nonetheless, when $U(T,k)$ has two quasienergy gaps 
$0\leq\epsilon<\epsilon'<2\pi/T$ then
$H^{\text{eff}}_{\epsilon'}(k)-H^{\text{eff}}_\epsilon(k)=\frac{2\pi}{T}
P_{\epsilon,\epsilon'}(k)$, where $P_{\epsilon,\epsilon'}(k)$
are projectors on states $\ket{\psi_j}$ with eigenvalues $\lambda_j$ such that $\epsilon < \arg(\lambda_j)/T < \epsilon'$. 
The first Chern number of the vector bundle $\mathcal{E}_{\epsilon,\epsilon'}$ on which $P_{\epsilon,\epsilon'}(k)$ 
projects is then related to indices $W$ by $c_1(\mathcal{E}_{\epsilon,\epsilon'}) = W_{\epsilon'}[U] - W_{\epsilon}[U]$ \cite{RudnerLindnerBergLevin2013}.

\textit{Index $K$ for TRI periodically driven systems.}~--
In static systems, time-reversal invariance is described at the level of Hamiltonian as $\TR H \TR^{-1} = H$
with the time-reversal operator $\TR = \ee^{- \ii \pi S_y/\hbar} \CC$, where $S$
is the spin operator and $\CC$ represents the complex conjugation.
More generally, a system is time-reversal invariant when $\TR H(t) \TR^{-1} = H(-t)$, or equivalently $\TR U(t) \TR^{-1} = U(-t)$, 
up to a choice of the origin of time.
For a family of Bloch Hamiltonians, this property translates into $\TR H(t,k) \TR^{-1} = H(-t,-k)$ i.e. $\TR U(t,k) \TR^{-1} = U(-t,-k)$ for the evolution operators.
\begin{figure}[h!]
\includegraphics{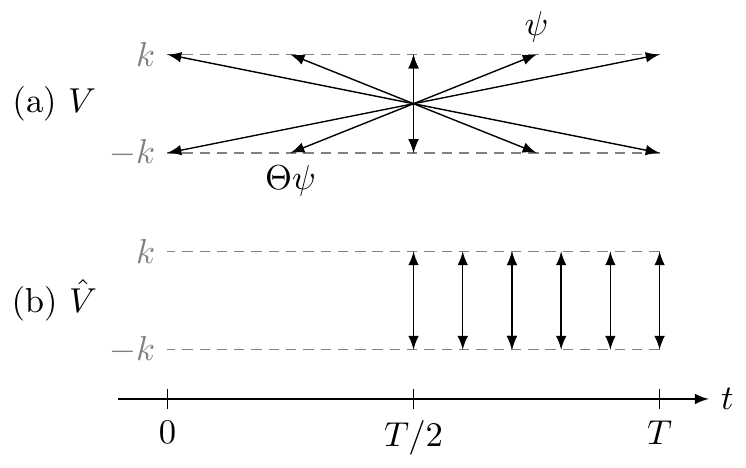}
\caption{\label{fig:figure_TR_constraint} 
Sketch of the periodized evolution over period $T$.
(a) Periodized evolution $V(t,k)$. Time-reversal relates pairs $(\psi, \Theta \psi)$ of states at $(t,k)$ and $(-t,-k)$, as shown by black arrows.
(b) Contracted half-evolution $\hat{V}(t,k)$. The second half of the initial evolution was discarded and replaced by a contraction respecting 
an equal-time constraint \eqref{eq:Vhat1b} depicted as black arrows.
}
\end{figure}

The first key observation is that, in a TRI periodically driven system, 
Kramers pairs $(\psi_j,\Theta\psi_j)$ of eigenstates of $U(T,k)$ and $U(T,-k)$,
respectively, \enquote{evolve in time} in opposite directions:
$\Theta U(t,k)\psi_j(k)=U(-t,-k)\Theta\psi_j(k)$ and similarly $\Theta V_\epsilon(t,k)\psi_j(k)=V_\epsilon(T-t,-k)\Theta\psi_j(k)$, as illustrated in 
Fig.~\ref{fig:figure_TR_constraint}(a). 
This property implies that $W_{\epsilon}[U] = 0$ because the contributions from 
two Kramers partners cancel. 
Consequently, also the first Chern numbers of quasienergy bands vanish in TRI periodically driven systems, as for the energy bands in the static case.
To circumvent this cancelation, we shall keep  the contribution only of one member of each Kramers pair by restricting the time evolution to times between
$t=0$ and $t=T/2$ (see Fig.~\ref{fig:figure_TR_constraint}(a)) in the same spirit 
than the construction by Moore and Balents in static systems~\cite{MooreBalents2007}.

The second key observation is that TR relates states at time $t=T/2$ to states at the same time.
We may thus directly compare the evolved Kramers partners at $t=T/2$.
Then it is possible to deform the map $k \mapsto V_{\epsilon}(T/2,k)$ to the identity
through a contraction of the Kramers pairs present at $t=T/2$.
This will result in a periodic map containing the information on the topological winding of 
Kramers pairs during the first half-period, see Fig.~\ref{fig:figure_TR_constraint}(b).

More concretely, we consider a smooth map $\widehat{V}_{\epsilon}$ 
from $[0,T]\times\BZ$ to $U(N)$ such that
\begin{subequations}
\label{eq:Vhat1} 
\begin{equation}
\label{eq:Vhat1a}
\widehat{V}_{\epsilon}(t,k) = V_{\epsilon}(t,k) \quad \quad \quad \quad  \text{for $0 \leq t \leq T/2$}
\end{equation}
and
\begin{equation}
\label{eq:Vhat1b}
\TR \widehat{V}_{\epsilon}(t,k) \TR^{-1} = \widehat{V}_{\epsilon}(t,-k)  \quad \ \text{for $T/2 \leq t \leq T$}
\end{equation}
\end{subequations} 
with $\widehat{V}_\epsilon(T,k)=I=\widehat{V}_\epsilon(0,k)$. 
The 
$\bm{Z}_2$-valued index $K$ is defined by the relation
\begin{equation}
	K_{\epsilon}[U] = \deg(\widehat{V}_{\epsilon})\ \;\text{mod}\ 2 
\label{eq:defK}
\end{equation}
(remember that $V_\epsilon$ is defined from $U$). This quantity is well-defined.
The existence of contractions \eqref{eq:Vhat1b}, the independence of 
$K_\epsilon[U]$ upon their choice, and the link with bundle Kane-Mele invariants (\ref{eq:KMK}) will be discussed elsewhere (see also Supplemental Materials \cite{our_supplemental}). In a semi-infinite system, $K_\epsilon[U]$ should give the parity 
of the number of Kramers pairs of edge states that lie in the corresponding bulk quasienergy gap.

Result \eqref{eq:KMK} greatly simplifies when spin is conserved.
In this case, the evolution operator $U$ is block-diagonal in the $(\uparrow,\downarrow)$ basis and
so is $V_{\epsilon}$. 
The two blocks are related by time-reversal and the $K$ index can be related to the~$W$ index 
of one of the spin blocks, namely
\begin{equation}
	\label{eq:K_spin_diagonal}
	K_{\epsilon}\left[\begin{pmatrix}
		U_{\uparrow} & 0 \\
		0 & U_{\downarrow}
	\end{pmatrix}\right] = \frac{W_{\epsilon}[U_{\uparrow}] - W_{\epsilon}[U_{\downarrow}]}{2} \quad \text{mod}\ 2 \ ,
\end{equation}
where $W_{\epsilon}[U_{\uparrow}] = W_{\epsilon}[\TR U_{\downarrow} \TR^{-1}] = - W_{\epsilon}[U_{\downarrow}]$.
This expression is reminiscent of the \enquote{spin Chern number}~\cite{ShengWengShengHaldane2006}. Indeed, when considering  the difference between indices at two quasienergy gaps, the usual spin Chern number is recovered.

\textit{Relation to the Wess-Zumino amplitude.}~-- 
The difference $K_{\epsilon'}[U]-K_\epsilon[U]$ is equal 
to the degree taken modulo 2 of the map $\widehat{V}_{\epsilon,\epsilon'}$ 
constructed as 
in \eqref{eq:Vhat1} from 
$V_{\epsilon,\epsilon'}=\ee^{-2\pi\ii t P_{\epsilon,\epsilon'}(k)/T}$. Due to
the relation $\Theta V_{\epsilon,\epsilon'}(t,k)\Theta^{-1}=
V_{\epsilon,\epsilon'}(t,-k)^{-1}$, the first 
half-period of time does not contribute to the integral 
for $\deg(\widehat{V}_{\epsilon,\epsilon'})$. 
The latter involves then only the contribution
of the contraction \eqref{eq:Vhat1b} of $V_{\epsilon,\epsilon'}(T/2)$.
Up to a factor, this contribution coincides with the Wess-Zumino action
\cite{PolyakovWiegmann1983,Witten1984} of the $U(N)$-valued field 
$V_{\epsilon,\epsilon'}(T/2,k)=I-2P_{\epsilon,\epsilon'}(k)$
defined on $\BZ$:
\begin{equation}
\deg(\widehat{V}_{\epsilon,\epsilon'})=
- \frac{1}{2\pi}
S_{\WZ}(V_{\epsilon,\epsilon'}(T/2)).
\end{equation}
The action $S_\WZ$ is normally determined modulo $2\pi$ but the condition 
\eqref{eq:Vhat1b}
for $\widehat{V}_{\epsilon,\epsilon'}$ makes it well-defined 
modulo~$4\pi$. We infer then from~\eqref{eq:KMK} the identity
\begin{equation}
(-1)^{\text{KM}(\mathcal{E}_{\epsilon,\epsilon'})} = 
\ee^{\frac{\ii}{2}S_\WZ(V_{\epsilon,\epsilon'}(T/2))}
\label{eq:KMWZ}
\end{equation}
relating the Kane-Mele invariant to the square root of the Wess-Zumino 
amplitude. This identity holds also in the static TRI case, 
providing a new expression for the 2D Kane-Mele invariant of the valence 
subbundle of states.  

\textit{Time-dependent lattice model.}~--
\begin{figure}[b!]
\includegraphics{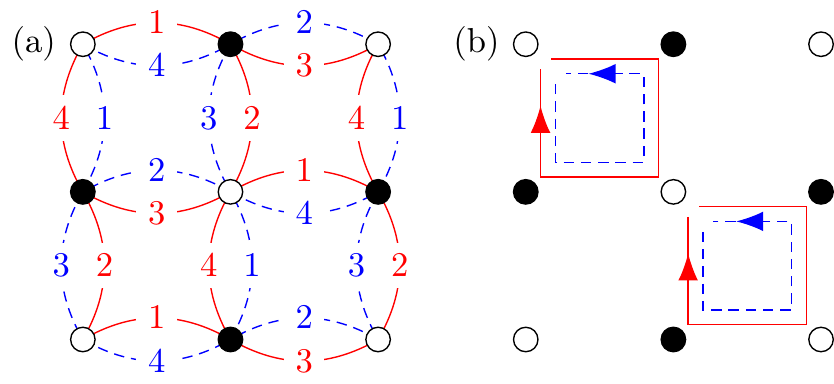}
\label{fig:schema_lattice_model}
\caption{Representation of the time evolution. 
(a) The only nonvanishing hopping amplitudes between sites are represented for each time step $\alpha = 1, \dots, 4$ as a link labeled with~$\alpha$. 
This is done for spin up (red solid lines) and spin down (blue dashed lines). 
(b) The time sequence of $U_{\alpha}$ is summarized around plaquettes, which mimics cyclotron orbits. Full and empty circles represent the two sublattices.
}
\label{fig:lattice}
\end{figure}
We shall now illustrate some of the basic properties
of TRI periodic evolution operators characterized by the
$K$ index  with a simple time-dependent lattice model.
A natural starting point is to define for spin $\uparrow$ states a periodic evolution possessing a nonvanishing invariant~$W_\epsilon$ 
inside a spectral gap, and to deduce by time-reversal symmetry the evolution of spin~$\downarrow$ states with the opposite~$W_\epsilon$.
Moreover, to verify that for a finite sample the index $K_\epsilon$
gives the parity of the number of pairs of edge states in
the corresponding gap independently of TRI gap-preserving perturbations, we add TRI spin flipping couplings to the model.
\begin{figure*}
\centering
 \includegraphics[width=1\textwidth]{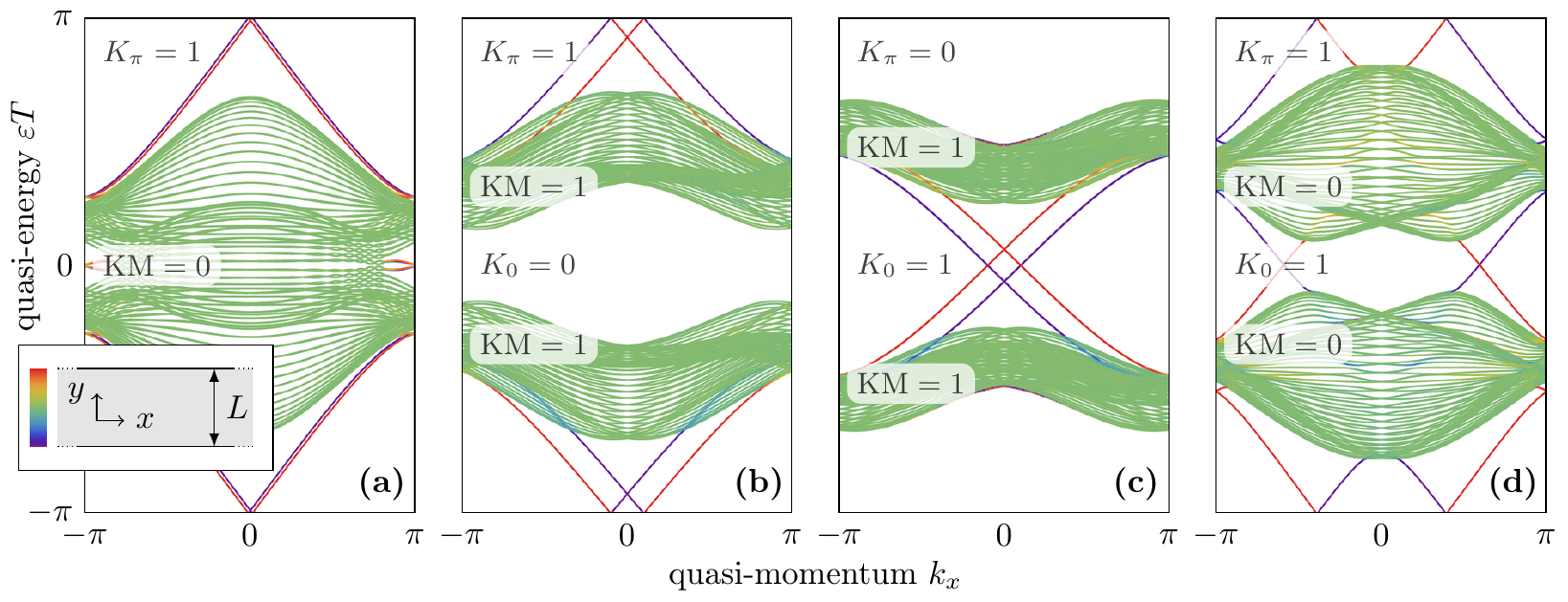}
 \caption{Quasienergy spectra of the TRI periodic evolution for a strip geometry (inset). The spectra reveal helical edge states in the gap $\epsilon=0$ or $\pi$.
 The parity of the number of pairs of edge states in a bulk gap 
localized on each boundary is given by the 
corresponding value $K_\epsilon$.
 Colors correspond to the density of states along $y$: red and purple states are localized at opposite edges (see inset). 
 The parameters are 
 (a) $J=3\pi$,			$J'=\pi$, 	$\Delta=0$,			
 (b) $J=3\pi/2$, 		$J'=1/2$, 	$\Delta=\pi/2$,		
 (c) $J=-5\pi$, 		$J'=1/2$, 	$\Delta=9\pi/2$,		
 (d) $J=15\pi/2$,		$J'=\pi$, 	$\Delta=2\pi$.
 For clarity, in the case (a), a small boundary mass term was added to distinguish edge states.
 }
 \label{fig:spectres}
\end{figure*}
For simplicity we define a time-step periodic dynamics
such that, during each step $\alpha$ of the evolution, the corresponding Hamiltonian $H_\alpha$ is constant in time.
One cycle of period $T$ is split into four steps $(\alpha-1)\frac{T}{4} \leq t < \alpha \frac{T}{4}$ so that the evolution operator
$U(T)=U_4U_3U_2U_1$ where $U_\alpha = \exp{(-\ii H_\alpha T/4)}$.
As the initial dynamics of spin~$\uparrow$, we consider a quantum analog of walks along classical cyclotron orbits on a square lattice  \cite{KitagawaBergRudnerDemler2010,RudnerLindnerBergLevin2013}. 
We distinguish two sublattices A and B and define the corresponding Hamiltonian for spin~$\uparrow$ for the first step $\alpha=1$ as
$ H_1^{\uparrow \uparrow} = J~ T_{+x}^{\text{A}\rightarrow \text{B}} + h.c.$
where $T^{A\to B}_{+x}$ is the translation operator by one horizontal lattice spacing from sublattice $A$ to $B$, see Fig.~\ref{fig:lattice}. 
Hamiltonians $H_{\alpha}^{\uparrow\uparrow}$ are obtained from $H_1^{\uparrow\uparrow}$ by replacing the translation operator $T^{A\to B}_{+x}$
by $T^{B\to A}_{-y}$ for $H_2^{\uparrow\uparrow}$,
by $T^{A\to B}_{-x}$ for $H_3^{\uparrow\uparrow}$, and
by $T^{B\to A}_{+y}$ for $H_4^{\uparrow\uparrow}$. 
Hamiltonians $H_\alpha^{\downarrow \downarrow}$ for spin $\downarrow$ states are deduced by TRI that imposes 
$H^{\downarrow \downarrow} (t,k)=\overline{H}^{\uparrow\uparrow}(-t,-k) $. They correspond to 
quantum evolution along orbits cycling in opposite direction.
In this initial model, the index $K_\epsilon$ associated to a spectral gap of $U(T)$ identifies modulo $2$ with the index $W_\epsilon$ of 
the evolution of the $\uparrow$ states, see \eqref{eq:K_spin_diagonal}. This is no longer the case when spin-flipping terms are added to the dynamics. 
They are incorporated into the model by adding to the Hamiltonian off-diagonal terms $H_\alpha^{\downarrow\uparrow}$
obtained from $H_\alpha^{\uparrow \uparrow}$ by replacing $J$ by $J'$, 
together with $H^{\uparrow \downarrow}(t,k) =- \overline{H}^{\downarrow \uparrow}(-t,-k)$, as imposed by the TRI constraint.  

The evolution operator over one period $U(T)$ is diagonalized for a strip geometry periodic in the $x$ direction. 
The quasienergies of the strip, together with the localization of the eigenstates from each edge, are shown in Fig.~\ref{fig:spectres}. 
Fig.~\ref{fig:spectres}(a) clearly shows in the only gap at $T\epsilon=\pi$ the existence of one pair of helical boundary states at each edge of the strip,
although the KM index associated with the unique band is necessarily zero.
The quasienergy gap does not close when $J'$ decreases to zero (see Supplemental Material \cite{our_supplemental}). 
Consequently, the value $K_\pi$ in the $\pi$ gap remains unchanged
and can be calculated from \eqref{eq:K_spin_diagonal}. We find $K_\pi=1$, in agreement with the number of edge states. 
 
Next, we consider a case with two bands i.e. with two gaps. 
To do so, we add a staggered on-site potential $\Delta$ and a fifth step $\alpha=5$ with $U_5=I$ \cite{RudnerLindnerBergLevin2013}.
Two typical situations are found and illustrate our formula \eqref{eq:KMK},
relating the Kane-Mele invariant associated to each quasienergy band and the index $K$ 
associated to $0$ and $\pi$ gaps. 
In Fig.~\ref{fig:spectres}(b) (respectively (c)) one pair of helical edge states appears in the 
$\pi$ gap (respectively $0$ gap).
We find in perfect agreement $K_0=0$ and  $K_\pi=1$ (Fig.~\ref{fig:spectres}(b))
and $K_0=1$, $K_\pi=0$ (Fig.~\ref{fig:spectres}(c))
and the Kane-Mele invariants are non-zero for each band. 
In contrast, Fig.~\ref{fig:spectres}(d) illustrates a situation where one pair of edge states lies in each gap, in agreement with
$K_0=1$ and $K_\pi=1$, so the Kane-Mele band invariants vanish.
This provides a TRI analog of the periodically-driven TR breaking phases characterized with zero
Chern band invariants but non-zero $W$ gap invariants \cite{RudnerLindnerBergLevin2013}.

\textit{Conclusions.}~--
We defined a new index characterizing topological properties of periodically driven 2D crystals constrained by time-reversal invariance, relating it to the Kane-Mele invariant 
of quasienergy bands for which we found a new expression in terms of the Wess-Zumino amplitude.
This paves the way for a study of physical properties associated to these topological states, such as the out-of-equilibrium 
DC transport \cite{Gu11, Kitagawa11, Kundu13} by the edge states that our analysis reveals. 
In particular, the search for experimental signatures in electronic systems
via transport properties in a multi-terminal geometry is a particularly interesting direction for future inverstigations,
but other proposals such as a local probe by means of tunneling current \cite{Fregoso14} are also conceivable.
Various physical setups, like anisotropic metamaterials \cite{Khanikaev13},
dielectric ring resonators~\cite{Hafezi11}, or shaken optical lattices \cite{Yan14}, seem to be good candidates to achieve experimentally such topological states
as long as dissipation is not too strong.
In particular, a Floquet analog of the Kane-Mele model seems to be achievable experimentally in a near future with cold atoms by
modulating simultaneously in time the trapping lattice and a magnetic field gradient. \cite{Jotzu14}.

\begin{acknowledgments}
We acknowledge Cl\'ement Tauber and Mickael Rechtsman for insightful discussions.
This work was supported by a grant from the Agence Nationale de la Recherche (ANR Blanc-2010 IsoTop).
\end{acknowledgments}

\bibliography{biblio}

\begin{thebibliography}{45}%
\makeatletter
\providecommand \@ifxundefined [1]{%
 \@ifx{#1\undefined}
}%
\providecommand \@ifnum [1]{%
 \ifnum #1\expandafter \@firstoftwo
 \else \expandafter \@secondoftwo
 \fi
}%
\providecommand \@ifx [1]{%
 \ifx #1\expandafter \@firstoftwo
 \else \expandafter \@secondoftwo
 \fi
}%
\providecommand \natexlab [1]{#1}%
\providecommand \enquote  [1]{``#1''}%
\providecommand \bibnamefont  [1]{#1}%
\providecommand \bibfnamefont [1]{#1}%
\providecommand \citenamefont [1]{#1}%
\providecommand \href@noop [0]{\@secondoftwo}%
\providecommand \href [0]{\begingroup \@sanitize@url \@href}%
\providecommand \@href[1]{\@@startlink{#1}\@@href}%
\providecommand \@@href[1]{\endgroup#1\@@endlink}%
\providecommand \@sanitize@url [0]{\catcode `\\12\catcode `\$12\catcode
  `\&12\catcode `\#12\catcode `\^12\catcode `\_12\catcode `\%12\relax}%
\providecommand \@@startlink[1]{}%
\providecommand \@@endlink[0]{}%
\providecommand \url  [0]{\begingroup\@sanitize@url \@url }%
\providecommand \@url [1]{\endgroup\@href {#1}{\urlprefix }}%
\providecommand \urlprefix  [0]{URL }%
\providecommand \Eprint [0]{\href }%
\providecommand \doibase [0]{http://dx.doi.org/}%
\providecommand \selectlanguage [0]{\@gobble}%
\providecommand \bibinfo  [0]{\@secondoftwo}%
\providecommand \bibfield  [0]{\@secondoftwo}%
\providecommand \translation [1]{[#1]}%
\providecommand \BibitemOpen [0]{}%
\providecommand \bibitemStop [0]{}%
\providecommand \bibitemNoStop [0]{.\EOS\space}%
\providecommand \EOS [0]{\spacefactor3000\relax}%
\providecommand \BibitemShut  [1]{\csname bibitem#1\endcsname}%
\let\auto@bib@innerbib\@empty
%</preamble>
\bibitem [{\citenamefont {Kane}\ and\ \citenamefont
  {Mele}(2005{\natexlab{a}})}]{KaneMele2005a}%
  \BibitemOpen
  \bibfield  {author} {\bibinfo {author} {\bibfnamefont {C.~L.}\ \bibnamefont
  {Kane}}\ and\ \bibinfo {author} {\bibfnamefont {E.~J.}\ \bibnamefont
  {Mele}},\ }\href {\doibase 10.1103/PhysRevLett.95.226801} {\bibfield
  {journal} {\bibinfo  {journal} {Phys. Rev. Lett.}\ }\textbf {\bibinfo
  {volume} {95}},\ \bibinfo {pages} {226801} (\bibinfo {year}
  {2005}{\natexlab{a}})},\ \Eprint {http://arxiv.org/abs/cond-mat/0411737}
  {cond-mat/0411737} \BibitemShut {NoStop}%
\bibitem [{\citenamefont {Kane}\ and\ \citenamefont
  {Mele}(2005{\natexlab{b}})}]{KaneMele2005}%
  \BibitemOpen
  \bibfield  {author} {\bibinfo {author} {\bibfnamefont {C.~L.}\ \bibnamefont
  {Kane}}\ and\ \bibinfo {author} {\bibfnamefont {E.~J.}\ \bibnamefont
  {Mele}},\ }\href {\doibase 10.1103/PhysRevLett.95.146802} {\bibfield
  {journal} {\bibinfo  {journal} {Phys. Rev. Lett.}\ }\textbf {\bibinfo
  {volume} {95}},\ \bibinfo {eid} {146802} (\bibinfo {year}
  {2005}{\natexlab{b}})},\ \Eprint {http://arxiv.org/abs/cond-mat/0506581}
  {cond-mat/0506581} \BibitemShut {NoStop}%
\bibitem [{\citenamefont {{Bernevig}}\ \emph {et~al.}(2006)\citenamefont
  {{Bernevig}}, \citenamefont {{Hughes}},\ and\ \citenamefont
  {{Zhang}}}]{BernevigHughesZhang2006}%
  \BibitemOpen
  \bibfield  {author} {\bibinfo {author} {\bibfnamefont {B.~A.}\ \bibnamefont
  {{Bernevig}}}, \bibinfo {author} {\bibfnamefont {T.~L.}\ \bibnamefont
  {{Hughes}}}, \ and\ \bibinfo {author} {\bibfnamefont {S.-C.}\ \bibnamefont
  {{Zhang}}},\ }\href {\doibase 10.1126/science.1133734} {\bibfield  {journal}
  {\bibinfo  {journal} {Science}\ }\textbf {\bibinfo {volume} {314}},\ \bibinfo
  {pages} {1757} (\bibinfo {year} {2006})},\ \Eprint
  {http://arxiv.org/abs/cond-mat/0611399} {cond-mat/0611399} \BibitemShut
  {NoStop}%
\bibitem [{\citenamefont {K{\"o}nig}\ \emph {et~al.}(2007)\citenamefont
  {K{\"o}nig}, \citenamefont {Wiedmann}, \citenamefont {Br{\"u}ne},
  \citenamefont {Roth}, \citenamefont {Buhmann}, \citenamefont {Molenkamp},
  \citenamefont {Qi},\ and\ \citenamefont {Zhang}}]{Konig2007}%
  \BibitemOpen
  \bibfield  {author} {\bibinfo {author} {\bibfnamefont {M.}~\bibnamefont
  {K{\"o}nig}}, \bibinfo {author} {\bibfnamefont {S.}~\bibnamefont {Wiedmann}},
  \bibinfo {author} {\bibfnamefont {C.}~\bibnamefont {Br{\"u}ne}}, \bibinfo
  {author} {\bibfnamefont {A.}~\bibnamefont {Roth}}, \bibinfo {author}
  {\bibfnamefont {H.}~\bibnamefont {Buhmann}}, \bibinfo {author} {\bibfnamefont
  {L.~W.}\ \bibnamefont {Molenkamp}}, \bibinfo {author} {\bibfnamefont {X.-L.}\
  \bibnamefont {Qi}}, \ and\ \bibinfo {author} {\bibfnamefont {S.-C.}\
  \bibnamefont {Zhang}},\ }\href {\doibase 10.1126/science.1148047} {\bibfield
  {journal} {\bibinfo  {journal} {Science}\ }\textbf {\bibinfo {volume}
  {318}},\ \bibinfo {pages} {766} (\bibinfo {year} {2007})},\ \Eprint
  {http://arxiv.org/abs/0710.0582} {0710.0582} \BibitemShut {NoStop}%
\bibitem [{\citenamefont {von Klitzing}\ \emph {et~al.}(1980)\citenamefont {von
  Klitzing}, \citenamefont {Dorda},\ and\ \citenamefont
  {Pepper}}]{KlitzingDordaPepper1980}%
  \BibitemOpen
  \bibfield  {author} {\bibinfo {author} {\bibfnamefont {K.}~\bibnamefont {von
  Klitzing}}, \bibinfo {author} {\bibfnamefont {G.}~\bibnamefont {Dorda}}, \
  and\ \bibinfo {author} {\bibfnamefont {M.}~\bibnamefont {Pepper}},\ }\href
  {\doibase 10.1103/PhysRevLett.45.494} {\bibfield  {journal} {\bibinfo
  {journal} {Phys. Rev. Lett.}\ }\textbf {\bibinfo {volume} {45}},\ \bibinfo
  {pages} {494} (\bibinfo {year} {1980})}\BibitemShut {NoStop}%
\bibitem [{\citenamefont {Thouless}\ \emph {et~al.}(1982)\citenamefont
  {Thouless}, \citenamefont {Kohmoto}, \citenamefont {Nightingale},\ and\
  \citenamefont {den Nijs}}]{ThoulessKohmotoNightingaleNijs1982}%
  \BibitemOpen
  \bibfield  {author} {\bibinfo {author} {\bibfnamefont {D.~J.}\ \bibnamefont
  {Thouless}}, \bibinfo {author} {\bibfnamefont {M.}~\bibnamefont {Kohmoto}},
  \bibinfo {author} {\bibfnamefont {M.~P.}\ \bibnamefont {Nightingale}}, \ and\
  \bibinfo {author} {\bibfnamefont {M.}~\bibnamefont {den Nijs}},\ }\href
  {\doibase 10.1103/PhysRevLett.49.405} {\bibfield  {journal} {\bibinfo
  {journal} {Phys. Rev. Lett.}\ }\textbf {\bibinfo {volume} {49}},\ \bibinfo
  {pages} {405} (\bibinfo {year} {1982})}\BibitemShut {NoStop}%
\bibitem [{\citenamefont {{Avron}}\ \emph {et~al.}(1983)\citenamefont
  {{Avron}}, \citenamefont {{Seiler}},\ and\ \citenamefont
  {{Simon}}}]{AvronSeilerSimon1983}%
  \BibitemOpen
  \bibfield  {author} {\bibinfo {author} {\bibfnamefont {J.~E.}\ \bibnamefont
  {{Avron}}}, \bibinfo {author} {\bibfnamefont {R.}~\bibnamefont {{Seiler}}}, \
  and\ \bibinfo {author} {\bibfnamefont {B.}~\bibnamefont {{Simon}}},\ }\href
  {\doibase 10.1103/PhysRevLett.51.51} {\bibfield  {journal} {\bibinfo
  {journal} {Phys. Rev. Lett.}\ }\textbf {\bibinfo {volume} {51}},\ \bibinfo
  {pages} {51} (\bibinfo {year} {1983})}\BibitemShut {NoStop}%
\bibitem [{\citenamefont {{Bellissard}}\ \emph {et~al.}(1994)\citenamefont
  {{Bellissard}}, \citenamefont {{van Elst}},\ and\ \citenamefont
  {{Schulz-Baldes}}}]{Bellissard1994}%
  \BibitemOpen
  \bibfield  {author} {\bibinfo {author} {\bibfnamefont {J.}~\bibnamefont
  {{Bellissard}}}, \bibinfo {author} {\bibfnamefont {A.}~\bibnamefont {{van
  Elst}}}, \ and\ \bibinfo {author} {\bibfnamefont {H.}~\bibnamefont
  {{Schulz-Baldes}}},\ }\href {\doibase 10.1063/1.530758} {\bibfield  {journal}
  {\bibinfo  {journal} {J. Math. Phys.}\ }\textbf {\bibinfo {volume} {35}},\
  \bibinfo {pages} {5373} (\bibinfo {year} {1994})},\ \Eprint
  {http://arxiv.org/abs/cond-mat/9411052} {cond-mat/9411052} \BibitemShut
  {NoStop}%
\bibitem [{\citenamefont {Laughlin}(1981)}]{Laughlin1981}%
  \BibitemOpen
  \bibfield  {author} {\bibinfo {author} {\bibfnamefont {R.~B.}\ \bibnamefont
  {Laughlin}},\ }\href {\doibase 10.1103/PhysRevB.23.5632} {\bibfield
  {journal} {\bibinfo  {journal} {Phys. Rev. B}\ }\textbf {\bibinfo {volume}
  {23}},\ \bibinfo {pages} {5632} (\bibinfo {year} {1981})}\BibitemShut
  {NoStop}%
\bibitem [{\citenamefont {Thouless}(1983)}]{Thouless1983}%
  \BibitemOpen
  \bibfield  {author} {\bibinfo {author} {\bibfnamefont {D.~J.}\ \bibnamefont
  {Thouless}},\ }\href {\doibase 10.1103/PhysRevB.27.6083} {\bibfield
  {journal} {\bibinfo  {journal} {Phys. Rev. B}\ }\textbf {\bibinfo {volume}
  {27}},\ \bibinfo {pages} {6083} (\bibinfo {year} {1983})}\BibitemShut
  {NoStop}%
\bibitem [{\citenamefont {Fu}\ and\ \citenamefont {Kane}(2007)}]{FuKane2007}%
  \BibitemOpen
  \bibfield  {author} {\bibinfo {author} {\bibfnamefont {L.}~\bibnamefont
  {Fu}}\ and\ \bibinfo {author} {\bibfnamefont {C.~L.}\ \bibnamefont {Kane}},\
  }\href {\doibase 10.1103/PhysRevB.76.045302} {\bibfield  {journal} {\bibinfo
  {journal} {Phys. Rev. B}\ }\textbf {\bibinfo {volume} {76}},\ \bibinfo {eid}
  {045302} (\bibinfo {year} {2007})},\ \Eprint
  {http://arxiv.org/abs/cond-mat/0611341} {cond-mat/0611341} \BibitemShut
  {NoStop}%
\bibitem [{\citenamefont {{Meidan}}\ \emph {et~al.}(2011)\citenamefont
  {{Meidan}}, \citenamefont {{Micklitz}},\ and\ \citenamefont
  {{Brouwer}}}]{MeidanMicklitzBrouwer2011}%
  \BibitemOpen
  \bibfield  {author} {\bibinfo {author} {\bibfnamefont {D.}~\bibnamefont
  {{Meidan}}}, \bibinfo {author} {\bibfnamefont {T.}~\bibnamefont
  {{Micklitz}}}, \ and\ \bibinfo {author} {\bibfnamefont {P.~W.}\ \bibnamefont
  {{Brouwer}}},\ }\href {\doibase 10.1103/PhysRevB.84.195410} {\bibfield
  {journal} {\bibinfo  {journal} {Phys. Rev. B}\ }\textbf {\bibinfo {volume}
  {84}},\ \bibinfo {pages} {195410} (\bibinfo {year} {2011})},\ \Eprint
  {http://arxiv.org/abs/1107.2215} {1107.2215} \BibitemShut {NoStop}%
\bibitem [{\citenamefont {Fulga}\ \emph {et~al.}(2012)\citenamefont {Fulga},
  \citenamefont {Hassler},\ and\ \citenamefont
  {Akhmerov}}]{FulgaHasslerAkhmerov2012}%
  \BibitemOpen
  \bibfield  {author} {\bibinfo {author} {\bibfnamefont {I.~C.}\ \bibnamefont
  {Fulga}}, \bibinfo {author} {\bibfnamefont {F.}~\bibnamefont {Hassler}}, \
  and\ \bibinfo {author} {\bibfnamefont {A.~R.}\ \bibnamefont {Akhmerov}},\
  }\href {\doibase 10.1103/PhysRevB.85.165409} {\bibfield  {journal} {\bibinfo
  {journal} {Phys. Rev. B}\ }\textbf {\bibinfo {volume} {85}},\ \bibinfo
  {pages} {165409} (\bibinfo {year} {2012})},\ \Eprint
  {http://arxiv.org/abs/1106.6351} {1106.6351} \BibitemShut {NoStop}%
\bibitem [{\citenamefont {Inoue}\ and\ \citenamefont {Tanaka}(2010)}]{Inoue10}%
  \BibitemOpen
  \bibfield  {author} {\bibinfo {author} {\bibfnamefont {J.-i.}\ \bibnamefont
  {Inoue}}\ and\ \bibinfo {author} {\bibfnamefont {A.}~\bibnamefont {Tanaka}},\
  }\href {\doibase 10.1103/PhysRevLett.105.017401} {\bibfield  {journal}
  {\bibinfo  {journal} {Phys. Rev. Lett.}\ }\textbf {\bibinfo {volume} {105}},\
  \bibinfo {pages} {017401} (\bibinfo {year} {2010})}\BibitemShut {NoStop}%
\bibitem [{\citenamefont {{Lindner}}\ \emph {et~al.}(2011)\citenamefont
  {{Lindner}}, \citenamefont {{Refael}},\ and\ \citenamefont
  {{Galitski}}}]{LindnerRefaelGalitski2011}%
  \BibitemOpen
  \bibfield  {author} {\bibinfo {author} {\bibfnamefont {N.~H.}\ \bibnamefont
  {{Lindner}}}, \bibinfo {author} {\bibfnamefont {G.}~\bibnamefont {{Refael}}},
  \ and\ \bibinfo {author} {\bibfnamefont {V.}~\bibnamefont {{Galitski}}},\
  }\href {\doibase 10.1038/nphys1926} {\bibfield  {journal} {\bibinfo
  {journal} {Nature Phys.}\ }\textbf {\bibinfo {volume} {7}},\ \bibinfo {pages}
  {490} (\bibinfo {year} {2011})},\ \Eprint {http://arxiv.org/abs/1008.1792}
  {1008.1792} \BibitemShut {NoStop}%
\bibitem [{\citenamefont {Kitagawa}\ \emph {et~al.}(2011)\citenamefont
  {Kitagawa}, \citenamefont {Oka}, \citenamefont {Brataas}, \citenamefont
  {Fu},\ and\ \citenamefont {Demler}}]{Kitagawa11}%
  \BibitemOpen
  \bibfield  {author} {\bibinfo {author} {\bibfnamefont {T.}~\bibnamefont
  {Kitagawa}}, \bibinfo {author} {\bibfnamefont {T.}~\bibnamefont {Oka}},
  \bibinfo {author} {\bibfnamefont {A.}~\bibnamefont {Brataas}}, \bibinfo
  {author} {\bibfnamefont {L.}~\bibnamefont {Fu}}, \ and\ \bibinfo {author}
  {\bibfnamefont {E.}~\bibnamefont {Demler}},\ }\href {\doibase
  10.1103/PhysRevB.84.235108} {\bibfield  {journal} {\bibinfo  {journal} {Phys.
  Rev. B}\ }\textbf {\bibinfo {volume} {84}},\ \bibinfo {pages} {235108}
  (\bibinfo {year} {2011})}\BibitemShut {NoStop}%
\bibitem [{\citenamefont {Kitagawa}\ \emph {et~al.}(2010)\citenamefont
  {Kitagawa}, \citenamefont {Berg}, \citenamefont {Rudner},\ and\ \citenamefont
  {Demler}}]{KitagawaBergRudnerDemler2010}%
  \BibitemOpen
  \bibfield  {author} {\bibinfo {author} {\bibfnamefont {T.}~\bibnamefont
  {Kitagawa}}, \bibinfo {author} {\bibfnamefont {E.}~\bibnamefont {Berg}},
  \bibinfo {author} {\bibfnamefont {M.}~\bibnamefont {Rudner}}, \ and\ \bibinfo
  {author} {\bibfnamefont {E.}~\bibnamefont {Demler}},\ }\href {\doibase
  10.1103/PhysRevB.82.235114} {\bibfield  {journal} {\bibinfo  {journal} {Phys.
  Rev. B}\ }\textbf {\bibinfo {volume} {82}},\ \bibinfo {pages} {235114}
  (\bibinfo {year} {2010})},\ \Eprint {http://arxiv.org/abs/1010.6126}
  {1010.6126} \BibitemShut {NoStop}%
\bibitem [{\citenamefont {Rudner}\ \emph {et~al.}(2013)\citenamefont {Rudner},
  \citenamefont {Lindner}, \citenamefont {Berg},\ and\ \citenamefont
  {Levin}}]{RudnerLindnerBergLevin2013}%
  \BibitemOpen
  \bibfield  {author} {\bibinfo {author} {\bibfnamefont {M.~S.}\ \bibnamefont
  {Rudner}}, \bibinfo {author} {\bibfnamefont {N.~H.}\ \bibnamefont {Lindner}},
  \bibinfo {author} {\bibfnamefont {E.}~\bibnamefont {Berg}}, \ and\ \bibinfo
  {author} {\bibfnamefont {M.}~\bibnamefont {Levin}},\ }\href {\doibase
  10.1103/PhysRevX.3.031005} {\bibfield  {journal} {\bibinfo  {journal} {Phys.
  Rev. X}\ }\textbf {\bibinfo {volume} {3}},\ \bibinfo {pages} {031005}
  (\bibinfo {year} {2013})},\ \Eprint {http://arxiv.org/abs/1212.3324}
  {1212.3324} \BibitemShut {NoStop}%
\bibitem [{\citenamefont {{Wang}}\ \emph {et~al.}(2013)\citenamefont {{Wang}},
  \citenamefont {{Steinberg}}, \citenamefont {{Jarillo-Herrero}},\ and\
  \citenamefont {{Gedik}}}]{Wang13}%
  \BibitemOpen
  \bibfield  {author} {\bibinfo {author} {\bibfnamefont {Y.~H.}\ \bibnamefont
  {{Wang}}}, \bibinfo {author} {\bibfnamefont {H.}~\bibnamefont {{Steinberg}}},
  \bibinfo {author} {\bibfnamefont {P.}~\bibnamefont {{Jarillo-Herrero}}}, \
  and\ \bibinfo {author} {\bibfnamefont {N.}~\bibnamefont {{Gedik}}},\ }\href
  {\doibase 10.1126/science.1239834} {\bibfield  {journal} {\bibinfo  {journal}
  {Science}\ }\textbf {\bibinfo {volume} {342}},\ \bibinfo {pages} {453}
  (\bibinfo {year} {2013})},\ \Eprint {http://arxiv.org/abs/1310.7563}
  {1310.7563} \BibitemShut {NoStop}%
\bibitem [{\citenamefont {{Onishi}}\ \emph {et~al.}(2014)\citenamefont
  {{Onishi}}, \citenamefont {{Ren}}, \citenamefont {{Novak}}, \citenamefont
  {{Segawa}}, \citenamefont {{Ando}},\ and\ \citenamefont
  {{Tanaka}}}]{Onishi14}%
  \BibitemOpen
  \bibfield  {author} {\bibinfo {author} {\bibfnamefont {Y.}~\bibnamefont
  {{Onishi}}}, \bibinfo {author} {\bibfnamefont {Z.}~\bibnamefont {{Ren}}},
  \bibinfo {author} {\bibfnamefont {M.}~\bibnamefont {{Novak}}}, \bibinfo
  {author} {\bibfnamefont {K.}~\bibnamefont {{Segawa}}}, \bibinfo {author}
  {\bibfnamefont {Y.}~\bibnamefont {{Ando}}}, \ and\ \bibinfo {author}
  {\bibfnamefont {K.}~\bibnamefont {{Tanaka}}},\ }\href@noop {} {\bibfield
  {journal} {\bibinfo  {journal} {ArXiv e-prints}\ } (\bibinfo {year}
  {2014})},\ \Eprint {http://arxiv.org/abs/1403.2492} {1403.2492} \BibitemShut
  {NoStop}%
\bibitem [{\citenamefont {Fang}\ \emph {et~al.}(2012)\citenamefont {Fang},
  \citenamefont {Yu},\ and\ \citenamefont {Fan}}]{Fang12}%
  \BibitemOpen
  \bibfield  {author} {\bibinfo {author} {\bibfnamefont {K.}~\bibnamefont
  {Fang}}, \bibinfo {author} {\bibfnamefont {Z.}~\bibnamefont {Yu}}, \ and\
  \bibinfo {author} {\bibfnamefont {S.}~\bibnamefont {Fan}},\ }\href {\doibase
  doi:10.1038/nphoton.2012.236} {\bibfield  {journal} {\bibinfo  {journal}
  {Nature Photon.}\ }\textbf {\bibinfo {volume} {6}},\ \bibinfo {pages} {782}
  (\bibinfo {year} {2012})}\BibitemShut {NoStop}%
\bibitem [{\citenamefont {Pasek}\ and\ \citenamefont {Chong}(2014)}]{Pasek14}%
  \BibitemOpen
  \bibfield  {author} {\bibinfo {author} {\bibfnamefont {M.}~\bibnamefont
  {Pasek}}\ and\ \bibinfo {author} {\bibfnamefont {Y.~D.}\ \bibnamefont
  {Chong}},\ }\href {\doibase 10.1103/PhysRevB.89.075113} {\bibfield  {journal}
  {\bibinfo  {journal} {Phys. Rev. B}\ }\textbf {\bibinfo {volume} {89}},\
  \bibinfo {pages} {075113} (\bibinfo {year} {2014})}\BibitemShut {NoStop}%
\bibitem [{\citenamefont {Karzig}\ \emph {et~al.}(2014)\citenamefont {Karzig},
  \citenamefont {Bardyn}, \citenamefont {Lindner},\ and\ \citenamefont
  {Refael}}]{Karzig14}%
  \BibitemOpen
  \bibfield  {author} {\bibinfo {author} {\bibfnamefont {T.}~\bibnamefont
  {Karzig}}, \bibinfo {author} {\bibfnamefont {C.-E.}\ \bibnamefont {Bardyn}},
  \bibinfo {author} {\bibfnamefont {N.}~\bibnamefont {Lindner}}, \ and\
  \bibinfo {author} {\bibfnamefont {G.}~\bibnamefont {Refael}},\ }\href@noop {}
  {\bibfield  {journal} {\bibinfo  {journal} {arXiv:1406.4156}\ } (\bibinfo
  {year} {2014})}\BibitemShut {NoStop}%
\bibitem [{\citenamefont {{Kitagawa}}\ \emph {et~al.}(2012)\citenamefont
  {{Kitagawa}}, \citenamefont {{Broome}}, \citenamefont {{Fedrizzi}},
  \citenamefont {{Rudner}}, \citenamefont {{Berg}}, \citenamefont {{Kassal}},
  \citenamefont {{Aspuru-Guzik}}, \citenamefont {{Demler}},\ and\ \citenamefont
  {{White}}}]{Kitagawa2012}%
  \BibitemOpen
  \bibfield  {author} {\bibinfo {author} {\bibfnamefont {T.}~\bibnamefont
  {{Kitagawa}}}, \bibinfo {author} {\bibfnamefont {M.~A.}\ \bibnamefont
  {{Broome}}}, \bibinfo {author} {\bibfnamefont {A.}~\bibnamefont
  {{Fedrizzi}}}, \bibinfo {author} {\bibfnamefont {M.~S.}\ \bibnamefont
  {{Rudner}}}, \bibinfo {author} {\bibfnamefont {E.}~\bibnamefont {{Berg}}},
  \bibinfo {author} {\bibfnamefont {I.}~\bibnamefont {{Kassal}}}, \bibinfo
  {author} {\bibfnamefont {A.}~\bibnamefont {{Aspuru-Guzik}}}, \bibinfo
  {author} {\bibfnamefont {E.}~\bibnamefont {{Demler}}}, \ and\ \bibinfo
  {author} {\bibfnamefont {A.~G.}\ \bibnamefont {{White}}},\ }\href@noop {}
  {\bibfield  {journal} {\bibinfo  {journal} {Nature Commun.}\ }\textbf
  {\bibinfo {volume} {3}} (\bibinfo {year} {2012})},\ \Eprint
  {http://arxiv.org/abs/1105.5334} {1105.5334} \BibitemShut {NoStop}%
\bibitem [{\citenamefont {Rechtsman}\ \emph {et~al.}(2013)\citenamefont
  {Rechtsman}, \citenamefont {Zeuner}, \citenamefont {Plotnik}, \citenamefont
  {Lumer}, \citenamefont {Podolsky}, \citenamefont {Dreisow}, \citenamefont
  {Nolte}, \citenamefont {Segev},\ and\ \citenamefont
  {Szameit}}]{Rechtsman2013}%
  \BibitemOpen
  \bibfield  {author} {\bibinfo {author} {\bibfnamefont {M.~C.}\ \bibnamefont
  {Rechtsman}}, \bibinfo {author} {\bibfnamefont {J.~M.}\ \bibnamefont
  {Zeuner}}, \bibinfo {author} {\bibfnamefont {Y.}~\bibnamefont {Plotnik}},
  \bibinfo {author} {\bibfnamefont {Y.}~\bibnamefont {Lumer}}, \bibinfo
  {author} {\bibfnamefont {D.}~\bibnamefont {Podolsky}}, \bibinfo {author}
  {\bibfnamefont {F.}~\bibnamefont {Dreisow}}, \bibinfo {author} {\bibfnamefont
  {S.}~\bibnamefont {Nolte}}, \bibinfo {author} {\bibfnamefont
  {M.}~\bibnamefont {Segev}}, \ and\ \bibinfo {author} {\bibfnamefont
  {A.}~\bibnamefont {Szameit}},\ }\href@noop {} {\bibfield  {journal} {\bibinfo
   {journal} {Nature}\ }\textbf {\bibinfo {volume} {496}},\ \bibinfo {pages}
  {196} (\bibinfo {year} {2013})},\ \Eprint {http://arxiv.org/abs/1212.3146}
  {1212.3146} \BibitemShut {NoStop}%
\bibitem [{\citenamefont {Hauke}\ \emph {et~al.}(2012)\citenamefont {Hauke},
  \citenamefont {Tieleman}, \citenamefont {Celi}, \citenamefont
  {{\"O}lschl{\"a}ger}, \citenamefont {Simonet}, \citenamefont {Struck},
  \citenamefont {Weinberg}, \citenamefont {Windpassinger}, \citenamefont
  {Sengstock}, \citenamefont {Lewenstein},\ and\ \citenamefont
  {Eckardt}}]{Hauke12}%
  \BibitemOpen
  \bibfield  {author} {\bibinfo {author} {\bibfnamefont {P.}~\bibnamefont
  {Hauke}}, \bibinfo {author} {\bibfnamefont {O.}~\bibnamefont {Tieleman}},
  \bibinfo {author} {\bibfnamefont {A.}~\bibnamefont {Celi}}, \bibinfo {author}
  {\bibfnamefont {C.}~\bibnamefont {{\"O}lschl{\"a}ger}}, \bibinfo {author}
  {\bibfnamefont {J.}~\bibnamefont {Simonet}}, \bibinfo {author} {\bibfnamefont
  {J.}~\bibnamefont {Struck}}, \bibinfo {author} {\bibfnamefont
  {M.}~\bibnamefont {Weinberg}}, \bibinfo {author} {\bibfnamefont
  {P.}~\bibnamefont {Windpassinger}}, \bibinfo {author} {\bibfnamefont
  {K.}~\bibnamefont {Sengstock}}, \bibinfo {author} {\bibfnamefont
  {M.}~\bibnamefont {Lewenstein}}, \ and\ \bibinfo {author} {\bibfnamefont
  {A.}~\bibnamefont {Eckardt}},\ }\href {\doibase
  10.1103/PhysRevLett.109.145301} {\bibfield  {journal} {\bibinfo  {journal}
  {Phys. Rev. Lett.}\ }\textbf {\bibinfo {volume} {109}},\ \bibinfo {pages}
  {145301} (\bibinfo {year} {2012})}\BibitemShut {NoStop}%
\bibitem [{\citenamefont {Zheng}\ and\ \citenamefont {Zhai}(2014)}]{Zheng14}%
  \BibitemOpen
  \bibfield  {author} {\bibinfo {author} {\bibfnamefont {W.}~\bibnamefont
  {Zheng}}\ and\ \bibinfo {author} {\bibfnamefont {H.}~\bibnamefont {Zhai}},\
  }\href {\doibase 10.1103/PhysRevA.89.061603} {\bibfield  {journal} {\bibinfo
  {journal} {Phys. Rev. A}\ }\textbf {\bibinfo {volume} {89}},\ \bibinfo
  {pages} {061603} (\bibinfo {year} {2014})}\BibitemShut {NoStop}%
\bibitem [{\citenamefont {Reichl}\ and\ \citenamefont
  {Mueller}(2014)}]{Reichl14}%
  \BibitemOpen
  \bibfield  {author} {\bibinfo {author} {\bibfnamefont {M.~D.}\ \bibnamefont
  {Reichl}}\ and\ \bibinfo {author} {\bibfnamefont {E.~J.}\ \bibnamefont
  {Mueller}},\ }\href {\doibase 10.1103/PhysRevA.89.063628} {\bibfield
  {journal} {\bibinfo  {journal} {Phys. Rev. A}\ }\textbf {\bibinfo {volume}
  {89}},\ \bibinfo {pages} {063628} (\bibinfo {year} {2014})}\BibitemShut
  {NoStop}%
\bibitem [{\citenamefont {Jotzu}\ \emph {et~al.}(2014)\citenamefont {Jotzu},
  \citenamefont {Messer}, \citenamefont {Desbuquois}, \citenamefont {Lebrat},
  \citenamefont {Uehlinger}, \citenamefont {Greif},\ and\ \citenamefont
  {Esslinger}}]{Jotzu14}%
  \BibitemOpen
  \bibfield  {author} {\bibinfo {author} {\bibfnamefont {G.}~\bibnamefont
  {Jotzu}}, \bibinfo {author} {\bibfnamefont {M.}~\bibnamefont {Messer}},
  \bibinfo {author} {\bibfnamefont {R.}~\bibnamefont {Desbuquois}}, \bibinfo
  {author} {\bibfnamefont {M.}~\bibnamefont {Lebrat}}, \bibinfo {author}
  {\bibfnamefont {T.}~\bibnamefont {Uehlinger}}, \bibinfo {author}
  {\bibfnamefont {D.}~\bibnamefont {Greif}}, \ and\ \bibinfo {author}
  {\bibfnamefont {T.}~\bibnamefont {Esslinger}},\ }\href@noop {} {\bibfield
  {journal} {\bibinfo  {journal} {Nature}\ }\textbf {\bibinfo {volume} {515}},\
  \bibinfo {pages} {237} (\bibinfo {year} {2014})}\BibitemShut {NoStop}%
\bibitem [{\citenamefont {Jiang}\ \emph {et~al.}(2011)\citenamefont {Jiang},
  \citenamefont {Kitagawa}, \citenamefont {Alicea}, \citenamefont {Akhmerov},
  \citenamefont {Pekker}, \citenamefont {Refael}, \citenamefont {Cirac},
  \citenamefont {Demler}, \citenamefont {Lukin},\ and\ \citenamefont
  {Zoller}}]{Jiang11}%
  \BibitemOpen
  \bibfield  {author} {\bibinfo {author} {\bibfnamefont {L.}~\bibnamefont
  {Jiang}}, \bibinfo {author} {\bibfnamefont {T.}~\bibnamefont {Kitagawa}},
  \bibinfo {author} {\bibfnamefont {J.}~\bibnamefont {Alicea}}, \bibinfo
  {author} {\bibfnamefont {A.~R.}\ \bibnamefont {Akhmerov}}, \bibinfo {author}
  {\bibfnamefont {D.}~\bibnamefont {Pekker}}, \bibinfo {author} {\bibfnamefont
  {G.}~\bibnamefont {Refael}}, \bibinfo {author} {\bibfnamefont {J.~I.}\
  \bibnamefont {Cirac}}, \bibinfo {author} {\bibfnamefont {E.}~\bibnamefont
  {Demler}}, \bibinfo {author} {\bibfnamefont {M.~D.}\ \bibnamefont {Lukin}}, \
  and\ \bibinfo {author} {\bibfnamefont {P.}~\bibnamefont {Zoller}},\ }\href
  {\doibase 10.1103/PhysRevLett.106.220402} {\bibfield  {journal} {\bibinfo
  {journal} {Phys. Rev. Lett.}\ }\textbf {\bibinfo {volume} {106}},\ \bibinfo
  {pages} {220402} (\bibinfo {year} {2011})}\BibitemShut {NoStop}%
\bibitem [{\citenamefont {Asb{\'o}th}(2012)}]{Asboth12}%
  \BibitemOpen
  \bibfield  {author} {\bibinfo {author} {\bibfnamefont {J.~K.}\ \bibnamefont
  {Asb{\'o}th}},\ }\href {\doibase 10.1103/PhysRevB.86.195414} {\bibfield
  {journal} {\bibinfo  {journal} {Phys. Rev. B}\ }\textbf {\bibinfo {volume}
  {86}},\ \bibinfo {pages} {195414} (\bibinfo {year} {2012})}\BibitemShut
  {NoStop}%
\bibitem [{\citenamefont {Tarasinski}\ \emph {et~al.}(2014)\citenamefont
  {Tarasinski}, \citenamefont {Asb{\'o}th},\ and\ \citenamefont
  {Dahlhaus}}]{Tarasinski14}%
  \BibitemOpen
  \bibfield  {author} {\bibinfo {author} {\bibfnamefont {B.}~\bibnamefont
  {Tarasinski}}, \bibinfo {author} {\bibfnamefont {J.~K.}\ \bibnamefont
  {Asb{\'o}th}}, \ and\ \bibinfo {author} {\bibfnamefont {J.~P.}\ \bibnamefont
  {Dahlhaus}},\ }\href {\doibase 10.1103/PhysRevA.89.042327} {\bibfield
  {journal} {\bibinfo  {journal} {Phys. Rev. A}\ }\textbf {\bibinfo {volume}
  {89}},\ \bibinfo {pages} {042327} (\bibinfo {year} {2014})}\BibitemShut
  {NoStop}%
\bibitem [{\citenamefont {Dubrovin}\ \emph {et~al.}(1985)\citenamefont
  {Dubrovin}, \citenamefont {Fomenko},\ and\ \citenamefont
  {Novikov}}]{DubrovinFomenkoNovikovII}%
  \BibitemOpen
  \bibfield  {author} {\bibinfo {author} {\bibfnamefont {B.}~\bibnamefont
  {Dubrovin}}, \bibinfo {author} {\bibfnamefont {A.}~\bibnamefont {Fomenko}}, \
  and\ \bibinfo {author} {\bibfnamefont {S.}~\bibnamefont {Novikov}},\
  }\href@noop {} {\emph {\bibinfo {title} {Modern Geometry - Methods and
  Applications: Part II: The Geometry and Topology of Manifolds}}}\ (\bibinfo
  {publisher} {Springer},\ \bibinfo {year} {1985})\BibitemShut {NoStop}%
\bibitem [{\citenamefont {Bott}\ and\ \citenamefont
  {Seeley}(1978)}]{BottSeeley1978}%
  \BibitemOpen
  \bibfield  {author} {\bibinfo {author} {\bibfnamefont {R.}~\bibnamefont
  {Bott}}\ and\ \bibinfo {author} {\bibfnamefont {R.}~\bibnamefont {Seeley}},\
  }\href {http://projecteuclid.org/euclid.cmp/1103904396} {\bibfield  {journal}
  {\bibinfo  {journal} {Commun. Math. Phys.}\ }\textbf {\bibinfo {volume}
  {62}},\ \bibinfo {pages} {235} (\bibinfo {year} {1978})}\BibitemShut
  {NoStop}%
\bibitem [{\citenamefont {{Moore}}\ and\ \citenamefont
  {{Balents}}(2007)}]{MooreBalents2007}%
  \BibitemOpen
  \bibfield  {author} {\bibinfo {author} {\bibfnamefont {J.~E.}\ \bibnamefont
  {{Moore}}}\ and\ \bibinfo {author} {\bibfnamefont {L.}~\bibnamefont
  {{Balents}}},\ }\href {\doibase 10.1103/PhysRevB.75.121306} {\bibfield
  {journal} {\bibinfo  {journal} {Phys. Rev. B}\ }\textbf {\bibinfo {volume}
  {75}},\ \bibinfo {eid} {121306} (\bibinfo {year} {2007})},\ \Eprint
  {http://arxiv.org/abs/cond-mat/0607314} {cond-mat/0607314} \BibitemShut
  {NoStop}%
\bibitem [{our()}]{our_supplemental}%
  \BibitemOpen
  \href@noop {} {}\bibinfo {note} {See Supplemental Material for details on the
  lattice model and for a sketch of the proof of the invariance of the index
  K.}\BibitemShut {Stop}%
\bibitem [{\citenamefont {{Sheng}}\ \emph {et~al.}(2006)\citenamefont
  {{Sheng}}, \citenamefont {{Weng}}, \citenamefont {{Sheng}},\ and\
  \citenamefont {{Haldane}}}]{ShengWengShengHaldane2006}%
  \BibitemOpen
  \bibfield  {author} {\bibinfo {author} {\bibfnamefont {D.~N.}\ \bibnamefont
  {{Sheng}}}, \bibinfo {author} {\bibfnamefont {Z.~Y.}\ \bibnamefont {{Weng}}},
  \bibinfo {author} {\bibfnamefont {L.}~\bibnamefont {{Sheng}}}, \ and\
  \bibinfo {author} {\bibfnamefont {F.~D.~M.}\ \bibnamefont {{Haldane}}},\
  }\href {\doibase 10.1103/PhysRevLett.97.036808} {\bibfield  {journal}
  {\bibinfo  {journal} {Phys. Rev. Lett.}\ }\textbf {\bibinfo {volume} {97}},\
  \bibinfo {pages} {036808} (\bibinfo {year} {2006})},\ \Eprint
  {http://arxiv.org/abs/cond-mat/0603054} {cond-mat/0603054} \BibitemShut
  {NoStop}%
\bibitem [{\citenamefont {Polyakov}\ and\ \citenamefont
  {Wiegmann}(1983)}]{PolyakovWiegmann1983}%
  \BibitemOpen
  \bibfield  {author} {\bibinfo {author} {\bibfnamefont {A.}~\bibnamefont
  {Polyakov}}\ and\ \bibinfo {author} {\bibfnamefont {P.~B.}\ \bibnamefont
  {Wiegmann}},\ }\href {\doibase 10.1016/0370-2693(83)91104-8} {\bibfield
  {journal} {\bibinfo  {journal} {Phys. Lett. B}\ }\textbf {\bibinfo {volume}
  {131}},\ \bibinfo {pages} {121} (\bibinfo {year} {1983})}\BibitemShut
  {NoStop}%
\bibitem [{\citenamefont {Witten}(1984)}]{Witten1984}%
  \BibitemOpen
  \bibfield  {author} {\bibinfo {author} {\bibfnamefont {E.}~\bibnamefont
  {Witten}},\ }\href {\doibase 10.1007/BF01215276} {\bibfield  {journal}
  {\bibinfo  {journal} {Commun. Math. Phys.}\ }\textbf {\bibinfo {volume}
  {92}},\ \bibinfo {pages} {455} (\bibinfo {year} {1984})}\BibitemShut
  {NoStop}%
\bibitem [{\citenamefont {Gu}\ \emph {et~al.}(2011)\citenamefont {Gu},
  \citenamefont {Fertig}, \citenamefont {Arovas},\ and\ \citenamefont
  {Auerbach}}]{Gu11}%
  \BibitemOpen
  \bibfield  {author} {\bibinfo {author} {\bibfnamefont {Z.}~\bibnamefont
  {Gu}}, \bibinfo {author} {\bibfnamefont {H.~A.}\ \bibnamefont {Fertig}},
  \bibinfo {author} {\bibfnamefont {D.~P.}\ \bibnamefont {Arovas}}, \ and\
  \bibinfo {author} {\bibfnamefont {A.}~\bibnamefont {Auerbach}},\ }\href
  {\doibase 10.1103/PhysRevLett.107.216601} {\bibfield  {journal} {\bibinfo
  {journal} {Phys. Rev. Lett.}\ }\textbf {\bibinfo {volume} {107}},\ \bibinfo
  {pages} {216601} (\bibinfo {year} {2011})}\BibitemShut {NoStop}%
\bibitem [{\citenamefont {Kundu}\ and\ \citenamefont
  {Seradjeh}(2013)}]{Kundu13}%
  \BibitemOpen
  \bibfield  {author} {\bibinfo {author} {\bibfnamefont {A.}~\bibnamefont
  {Kundu}}\ and\ \bibinfo {author} {\bibfnamefont {B.}~\bibnamefont
  {Seradjeh}},\ }\href {\doibase 10.1103/PhysRevLett.111.136402} {\bibfield
  {journal} {\bibinfo  {journal} {Phys. Rev. Lett.}\ }\textbf {\bibinfo
  {volume} {111}},\ \bibinfo {pages} {136402} (\bibinfo {year}
  {2013})}\BibitemShut {NoStop}%
\bibitem [{\citenamefont {Fregoso}\ \emph {et~al.}(2014)\citenamefont
  {Fregoso}, \citenamefont {Dahlhaus},\ and\ \citenamefont
  {Moore}}]{Fregoso14}%
  \BibitemOpen
  \bibfield  {author} {\bibinfo {author} {\bibfnamefont {B.~M.}\ \bibnamefont
  {Fregoso}}, \bibinfo {author} {\bibfnamefont {J.~P.}\ \bibnamefont
  {Dahlhaus}}, \ and\ \bibinfo {author} {\bibfnamefont {J.~E.}\ \bibnamefont
  {Moore}},\ }\href {\doibase 10.1103/PhysRevB.90.155127} {\bibfield  {journal}
  {\bibinfo  {journal} {Phys. Rev. B}\ }\textbf {\bibinfo {volume} {90}},\
  \bibinfo {pages} {155127} (\bibinfo {year} {2014})}\BibitemShut {NoStop}%
\bibitem [{\citenamefont {{Khanikaev}}\ \emph {et~al.}(2013)\citenamefont
  {{Khanikaev}}, \citenamefont {{Hossein Mousavi}}, \citenamefont {{Tse}},
  \citenamefont {{Kargarian}}, \citenamefont {{MacDonald}},\ and\ \citenamefont
  {{Shvets}}}]{Khanikaev13}%
  \BibitemOpen
  \bibfield  {author} {\bibinfo {author} {\bibfnamefont {A.~B.}\ \bibnamefont
  {{Khanikaev}}}, \bibinfo {author} {\bibfnamefont {S.}~\bibnamefont {{Hossein
  Mousavi}}}, \bibinfo {author} {\bibfnamefont {W.-K.}\ \bibnamefont {{Tse}}},
  \bibinfo {author} {\bibfnamefont {M.}~\bibnamefont {{Kargarian}}}, \bibinfo
  {author} {\bibfnamefont {A.~H.}\ \bibnamefont {{MacDonald}}}, \ and\ \bibinfo
  {author} {\bibfnamefont {G.}~\bibnamefont {{Shvets}}},\ }\href {\doibase
  10.1038/nmat3520} {\bibfield  {journal} {\bibinfo  {journal} {Nature Mater.}\
  }\textbf {\bibinfo {volume} {12}},\ \bibinfo {pages} {233} (\bibinfo {year}
  {2013})},\ \Eprint {http://arxiv.org/abs/1204.5700} {1204.5700} \BibitemShut
  {NoStop}%
\bibitem [{\citenamefont {{Hafezi}}\ \emph {et~al.}(2011)\citenamefont
  {{Hafezi}}, \citenamefont {{Demler}}, \citenamefont {{Lukin}},\ and\
  \citenamefont {{Taylor}}}]{Hafezi11}%
  \BibitemOpen
  \bibfield  {author} {\bibinfo {author} {\bibfnamefont {M.}~\bibnamefont
  {{Hafezi}}}, \bibinfo {author} {\bibfnamefont {E.~A.}\ \bibnamefont
  {{Demler}}}, \bibinfo {author} {\bibfnamefont {M.~D.}\ \bibnamefont
  {{Lukin}}}, \ and\ \bibinfo {author} {\bibfnamefont {J.~M.}\ \bibnamefont
  {{Taylor}}},\ }\href {\doibase 10.1038/nphys2063} {\bibfield  {journal}
  {\bibinfo  {journal} {Nature Phys.}\ }\textbf {\bibinfo {volume} {7}},\
  \bibinfo {pages} {907} (\bibinfo {year} {2011})},\ \Eprint
  {http://arxiv.org/abs/1102.3256} {1102.3256} \BibitemShut {NoStop}%
\bibitem [{\citenamefont {{Yan}}\ \emph {et~al.}(2014)\citenamefont {{Yan}},
  \citenamefont {{Li}}, \citenamefont {{Yang}},\ and\ \citenamefont
  {{Wan}}}]{Yan14}%
  \BibitemOpen
  \bibfield  {author} {\bibinfo {author} {\bibfnamefont {Z.}~\bibnamefont
  {{Yan}}}, \bibinfo {author} {\bibfnamefont {B.}~\bibnamefont {{Li}}},
  \bibinfo {author} {\bibfnamefont {X.}~\bibnamefont {{Yang}}}, \ and\ \bibinfo
  {author} {\bibfnamefont {S.}~\bibnamefont {{Wan}}},\ }\href@noop {}
  {\bibfield  {journal} {\bibinfo  {journal} {ArXiv e-prints}\ } (\bibinfo
  {year} {2014})},\ \Eprint {http://arxiv.org/abs/1406.5087} {1406.5087}
  \BibitemShut {NoStop}%
\end{thebibliography}%

\end{document}